\documentclass[twocolumn,showpacs,aps,prl,superscriptaddress,unsortedaddress]{revtex4}

\usepackage{graphicx}
\usepackage{epsf}
\newcommand{\etal}{{\it et al.}}

\begin{document}

\title{Insensitivity of the superconducting gap to variation in T$_c$ in Zn-substituted Bi2212}
\author{Y. Lubashevsky}
\affiliation{Department of Physics, Technion, Haifa 32000, Israel}
\author{A. Garg}
\affiliation{Department of Physics, Technion, Haifa 32000, Israel}
\author{Y. Sassa}
\affiliation{Paul Scherrer Institute, CH-5232 Villigen PSI, Switzerland}
\author{M. Shi}
\affiliation{Swiss Light Source, Paul Scherrer Institute, CH-5232 Villigen PSI, Switzerland}
\author{A. Kanigel}
\affiliation{Department of Physics, Technion, Haifa 32000, Israel}

\date{\today}

\begin{abstract}
 The phase diagram of the superconducting high-Tc cuprates is governed by two energy scales: T$^{*}$, the temperature below which a gap is opened in the excitation spectrum, and T$_c$, the superconducting transition temperature. The way these two energy scales are reflected in the low-temperature energy gap is being intensively debated. Using Zn substitution and carefully controlled annealing we prepared a set of samples having the same T* but different T$_c$'s, and measured their gap using Angle Resolved Photoemission Spectroscopy (ARPES). We show that T$_c$ is not related to the gap shape or size, but it controls the size of the coherence peak at the gap edge. 
\end{abstract}
\maketitle

The relation between the pseudo-gap (PG) and superconductivity (SC) in the high-temperature superconducting cuprates (HTSC) remains a matter of controversy.  In fact, many believe that therein lies the key to understanding the cuprates' physics. 
During the last years a large body of experimental data was gathered but the results are inconclusive. While there is support for the concept of the PG being a state in which Cooper pairs are pre-formed \cite{pre-formed}, there are also numerous experiments providing evidence for a competing order \cite{competing_order}. 

A simple, but potentially very powerful, approach would be to look for correlations between measurable quantities and the two obvious energy scales that govern the cuprate phase diagram: T$_c$ and T$^{*}$. This approach is complicated by the fact that the parameter which sample growers control, the doping level of the sample, affects both T$_c$ and T$^{*}$ simultaneously. 
Here we overcome this problem using Zinc (Zn) substitution.  Zn substitution allows us to manipulate T$_c$ and T$^{*}$ separately.  We measure and compare the superconducting gaps of different samples, in that way we can observe directly the correlation between T$_c$, T$^{*}$ and the most basic property of a superconductor, its energy gap.   

\begin{figure}
\begin{center}
\includegraphics[width=8.25cm]{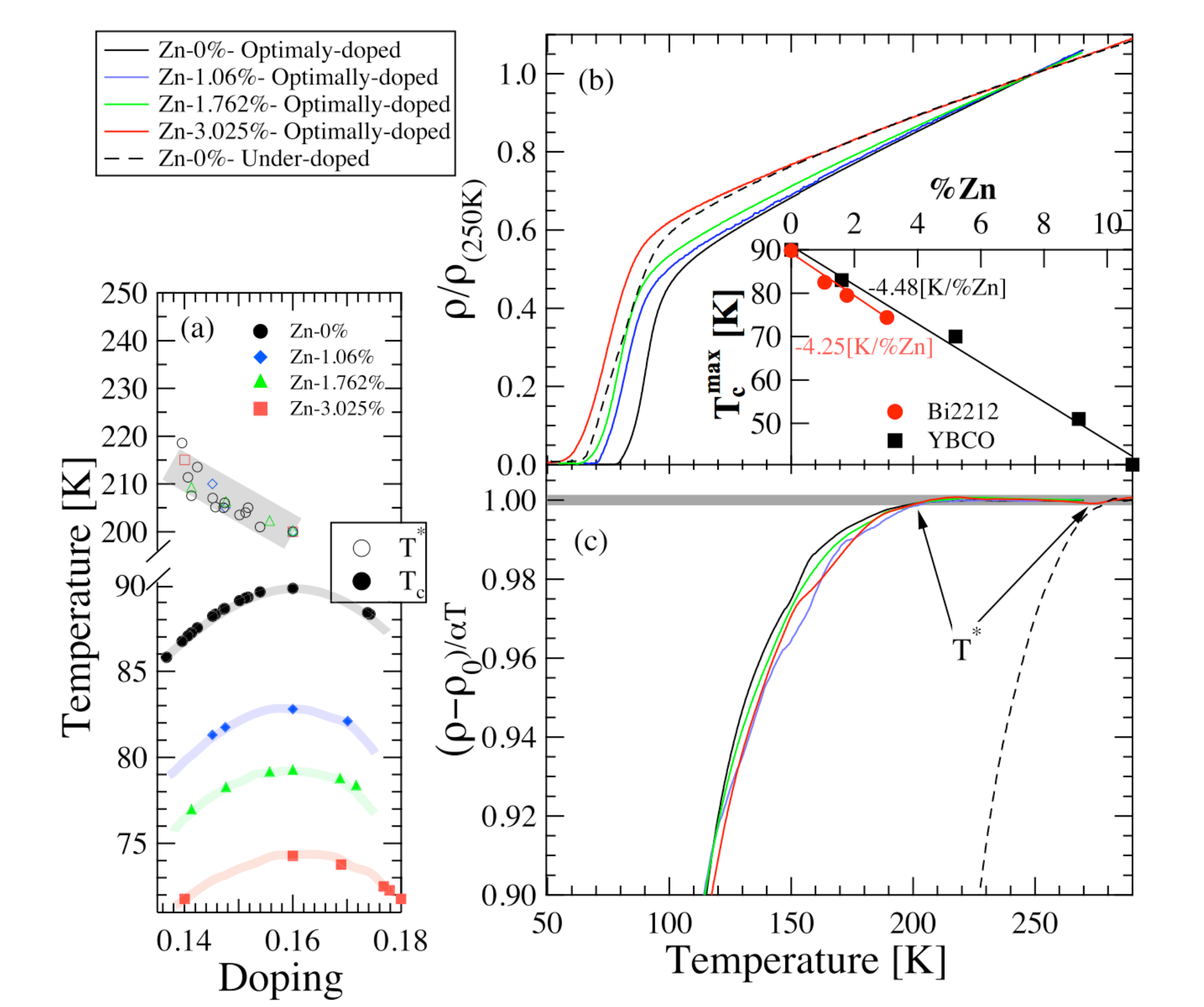}
\end{center}
\caption{
(a) Phase diagram of Zn-Bi2212. The Zn substitution creates parallel domes of T$_c$ vs doping. The maximal T$_c$ of each dome depends on the Zn amount, in contrast T$^*$ remains unchanged.  
(b) Resistivity as function of temperature for four optimally doped samples with different Zn substitution levels, normalized by the resistivity at $T=250K$. T$_c$ decreases with the addition of Zn. The dashed line is the resistivity of a pristine underdoped sample. Inset: T$_c$ as function of the Zn substitution level for Bi2212 and YBCO\cite{Chien}. 
(c) Deviation from linearity of the resistivity. The parameter $\alpha$ is the slope  of the high temperature linear part, and $\rho_0$ is the residual resistivity. All Zn-Bi2212 have the same T$^*$, but the UD sample has a higher T$^*$. 
}
\label{Fig1}
\end{figure}

Zn replaces Cu in the $CuO_2$ planes \cite{XiaoPRL}, without changing the carrier concentration.  
s-wave SCs are immune to scattering caused by non-magnetic impurities \cite{Anderson}, but because of the d-wave symmetry of the order parameter in the cuprates, Zn can reduce T$_c$ very effectively, although it is non-magnetic\cite{Alloul_Rev}. 
The rate at which Zn reduces T$_c$, $dT_c / d(Zn_{atom} \%)$, is different for different systems, ranging from about $-4.5$K for $YBa_2Cu_3O_{\delta}$ \cite{Chien} to $-9$K for $L_{2-x}Sr_xCuO_4$ \cite{Sanvik}(see the inset in Figure \ref{Fig1}b).

$\mu$SR experiments have shown that Zn substitution reduces the superfluid density ($n_s$) \cite{Nachumi, Nidermayer}; interestingly, the reduced $n_s$ and the lower T$_c$ ensures that the  Uemura relation \cite{uemura} remains intact.  The question of how Zn reduces the $n_s$ remains open; it has been suggested that SC is excluded  from a volume around each Zn atom, leaving a "swiss cheese"-like medium for SC \cite{Nachumi}. Others suggested that Zn, being a strong scatterer,  induces pair-breaking \cite{Nidermayer}.

For the purpose of this experiment we grew thin films of  Zn substituted Bi$_2$Sr$_2$CaCu$_2$O$_{8+\delta}$ (Zn-Bi2212) on LaAlO$_3$ substrates using  DC sputtering.  
The solubility limit of the Zn in the films is 3\%. Higher Zn concentration produced spurious phases in the samples. 
The doping level was controlled by annealing the films at low oxygen pressure. We used Wavelength-Dispersive x-ray Spectroscopy (WDS)  to measure the samples' composition. 
The results indicate that the films  are uniform and have the correct stoichiometry. For each Zn concentration a number of different samples were grown, with different doping levels from slightly underdoped to slightly overdoped, as shown in the phase diagram in Fig \ref{Fig1}(a).
The Zn substitution produces parallel T$_c$ vs. doping domes, where the doping is set by the oxygen amount.  The T$_c$s are from resistivity measurements and the doping level was calculated using the Presland formula \cite{presland}, with the appropriate T$_c^{max}$ for each Zn concentration. In contrast to the noticeable change in T$_c$ the Zn seems to have no effect on T$^{*}$. This is in agreement with the results of various other experiments \cite{JulienPRL, FongPRL, JanossyPRL}.

 Four optimally-doped samples were chosen for the ARPES experiment, with different  Zn/Cu ratios: Zn-0\% pristine Bi2212, Zn-1\% with 1.06(15)\%, Zn-2\% with 1.762(61)\% and Zn-3\% with 3.025(75)\%. In Fig \ref{Fig1}(b) the resistance as function of temperature for these samples is shown.  T$_c$ varies from 90K for the pristine sample to 74K for the 3\% sample.  We also present in the same figure an underdoped pristine Bi2212 sample, with T$_c=$77K. 

The pseudogap temperature was extracted from the resistance data by measuring the temperature at which the resistance starts to deviate from the linear behavior found at high temperature \cite{Raffy}. For each sample, we measured the high temperature slope, $\alpha$, and the residual resistivity, $\rho_0$. In Fig~\ref{Fig1}(c), we plot ($\rho(T)-\rho_0)/\alpha T$; T$^{*}$ is the point at which the value of ($\rho(T)-\rho_0)/\alpha T$ decreases below one. We find all the optimally doped samples to have the same $T^\star$ of 200K, while the UD sample has a higher T$^{*}$ of 270K. 
By manipulating the Zn concentration and the oxygen level it was possible to prepare a series of samples having the same T$^{*}$ but different T$_c$'s and two samples having similar T$_c$ but completely different T$^{*}$. 

Over the years ARPES has proven to be one of the most useful techniques for studying the electronic structure of the cuprate HTSC \cite{reviews}. Here we used ARPES to measure the momentum dependence of the gap in the excitations spectrum at low temperature. The ARPES measurements were done at the SIS beam line at the SLS, PSI, Switzerland, and at the NIM1 beam line at the SRC, Madison USA. All data were obtained using a Scienta R4000 analyzer using  22eV photons. To measure the SC gap, we took 12-15 momentum scans parallel to the  $\Gamma$-M direction.
For each cut we follow the peak-position in the Energy Distribution Curve (EDC). k$_F$ is defined as the point at which the separation between the peak and the chemical potential is minimal.
The peak-position of the EDC at k$_F$, after division by a resolution-broaden Fermi-function, defines the gap size at that point. E$_F$ is found by measuring the density of states of a gold layer evaporated on the sample holder. 

 \begin{figure}
\begin{center}
\includegraphics[width=8.0cm]{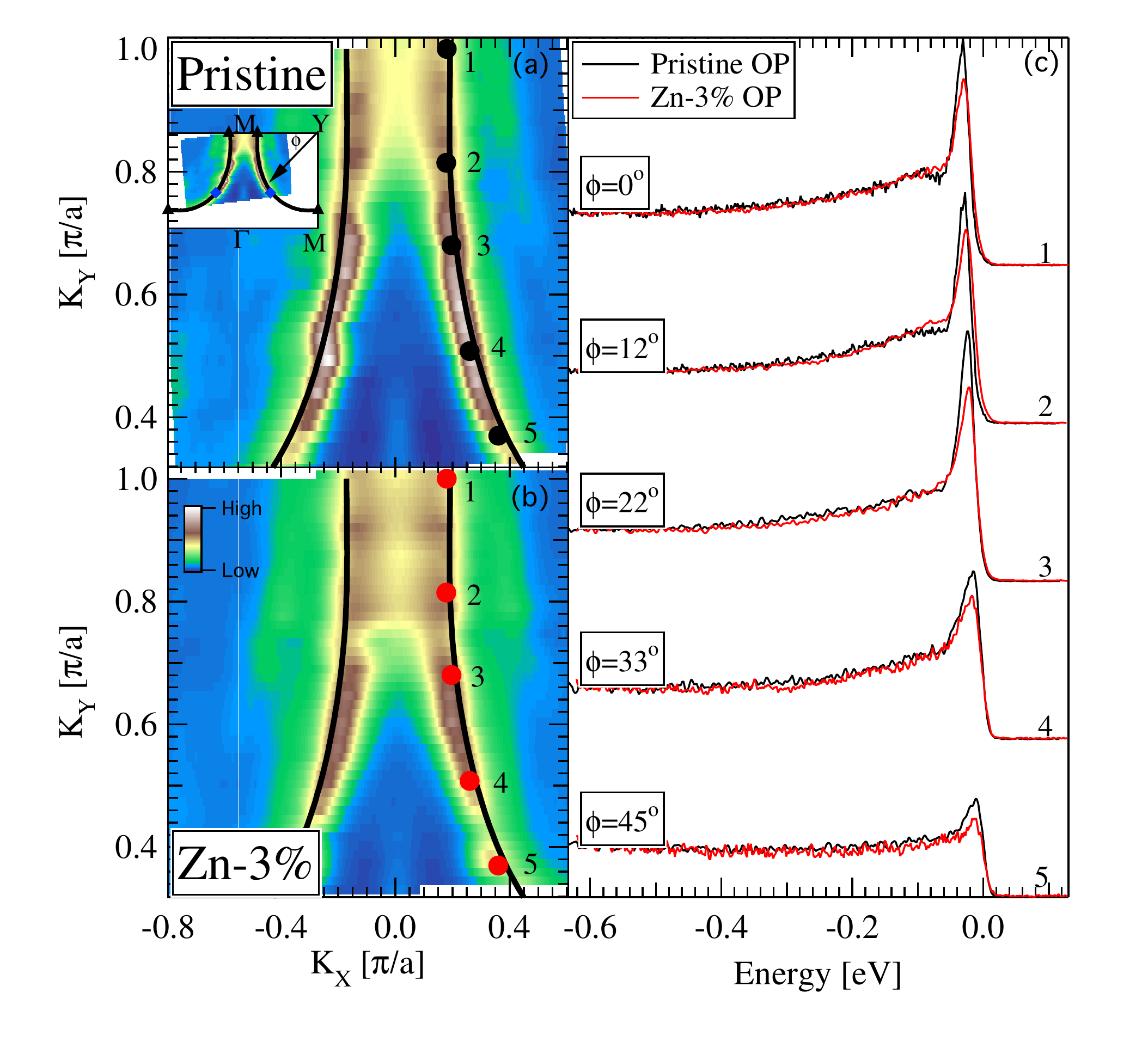}
\end{center}
\caption{ (a) and (b) ARPES intensity map at E$_F$ for the pristine and 3\% sample, respectively . The data is integrated over 40meV to reveal the underlying Fermi-surface. The solid lines represent the tight-binding Fermi-surface for optimally doped Bi2212. Both samples have the same Fermi-surface area indicating that they have the same carrier concentration. The inset shows the momentum range covered in panels a and b. (c) EDCs at several k$_F$ points (Represented in panels a and b by the black and red dots) for the pristine (black) and 3 \% Zn ( red) sample.
}
\label{Fig3}
\end{figure} 

The ARPES intensity maps at the Fermi energy of the pristine and Zn-3\% samples are shown in panels a and b of figure \ref{Fig3}. The black solid lines represent the tight-binding Fermi-surface of optimally-doped Bi2212 \cite{Mike_TB}. 
 In agreement with previous results, we do not find any change in the Fermi surface volume when Zn is added \cite{ Terashima2004,Zabolotnyy}, an indication that the Zn does not change the doping level. The Energy Distribution Curves (EDC) at the momentum points that are marked in panels a and b are shown in panel c. Well defined peaks can be observed in all the EDCs. Two conclusions can be drawn from figure \ref{Fig3}c: the height of the peaks is smaller in the Zn substituted sample, but the peak position remains the same indicating that there is no change in the gap. 

\begin{figure}
\begin{center}
\includegraphics[width=7.cm]{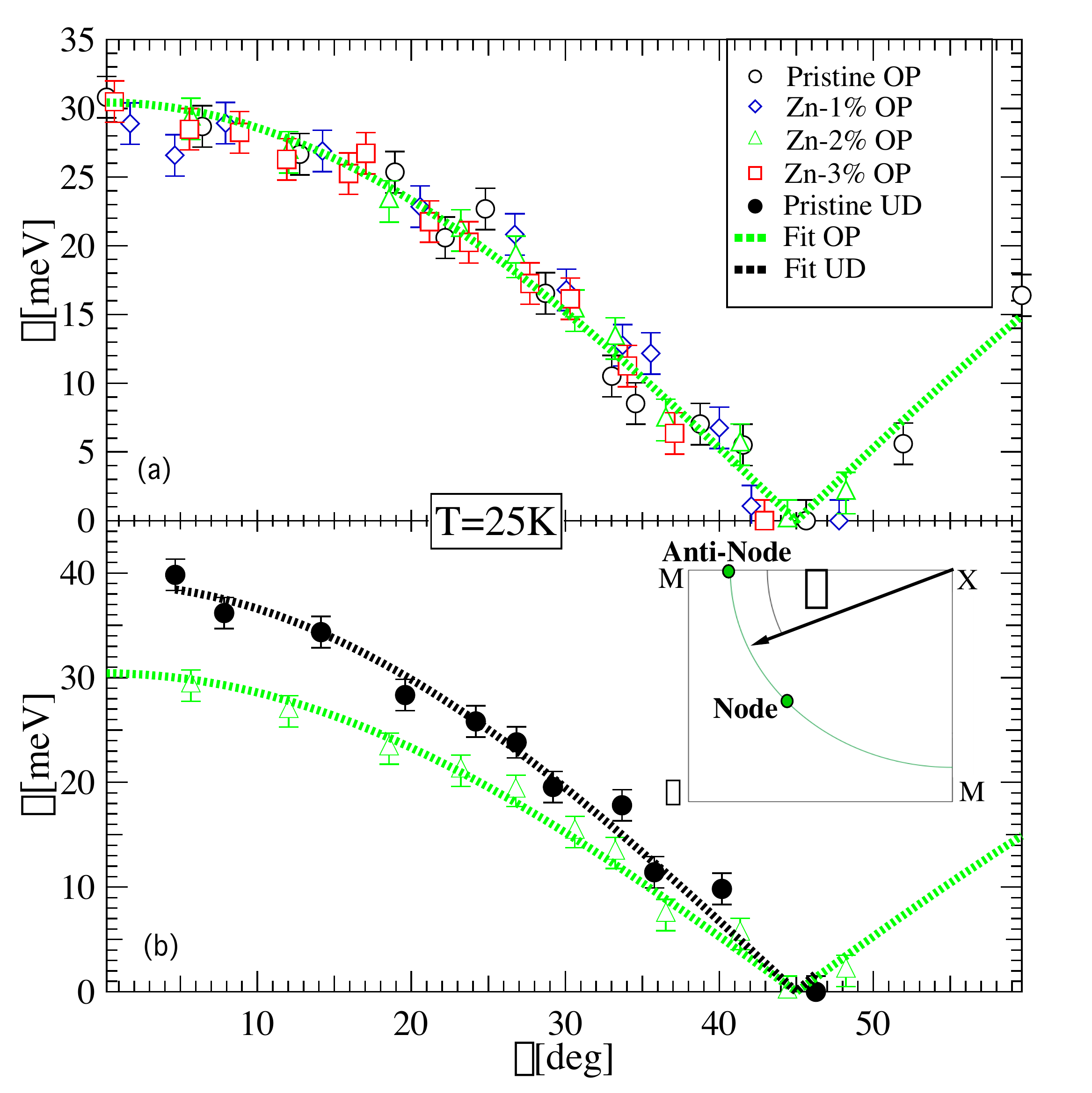}
\end{center}
\caption{
 (a) The momentum dependence of the superconducting gap of all Zn-Bi2212 samples at 25K. No   dependence on the Zn amount is observed, and the same d-wave function fits all data.
(b) The momentum dependence of the superconducting gap of two samples with similar T$_c$s: The pristine UD sample and the Zn-2\%. The lines are fits to the data using the simple d-wave function. A clear difference in gap can be seen. (Inset) The Brillouin zone of Bi2212 and the definition of the Fermi-surfce angle $\phi$. }
\label{Fig2}
\end{figure}

In Fig. \ref{Fig2}(a) the momentum dependence of the energy gap at T=25K of the four optimally doped samples is shown. The gaps are plotted as a function of the Fermi-surface angle, which is defined in the inset. The energy-gap of all the samples is identical although their T$_c$'s differ by up to 20\%. We emphasize that the gaps are identical over the entire Fermi surface, both in the nodal and anti-nodal regions.
 This is a remarkable result: it contradicts everything we know about Bardeen-Cooper-Schrieffer (BCS) SCs and about the way impurities reduce T$_c$ in these SCs. 
 
On the other hand our results reinforce the viewpoint that in the cuprates the gap is proportional to T$^*$ \cite{campuzano1999}; we kept T$^*$ constant and so the gap remains unchanged. In Fig \ref{Fig2}(b) we compare the gap-shape of the optimally doped samples (all having the same T$^{*}$) with that of the underdoped sample where T$^*$ is higher by 70K and where T$_c$ is similar to the T$_c$ of the Zn-2\% sample. The underdoped sample with the higher T$^*$ has a larger gap, as expected.

Recently, a number of groups reported gap-shapes that deviate substantially from the simple d-wave shape ( $\Delta = \Delta(0) cos (2 \phi)$) in various cuprate systems \cite{Tanaka2006,Lee2007,TerashimaLSCO,Kaminski2009}. According to these observations, there are two gaps covering different regions of the Fermi surface. The real SC gap opens in the nodal region, with a gap size proportional to T$_c$, while around the anti-nodal region there is a second gap, the Pseudogap, this gap-size follows T$^*$ as in the older data.
We, on the contrary, could fit all the data using a single simple d-wave gap function. The fit is represented by the dashed lines in Fig \ref{Fig2}. 
If one needs to choose a border-line in k-space that separates the two gaps, then a natural choice would be the Fermi-arc tip.
The Fermi-arcs are segments of the Fermi-surface, centered around the nodes, that are gapless above T$_c$ and are gapped-out below T$_c$ (apart for the node itself), as expected for a superconductor. The rest of the Fermi-surface remains gaped above T$_c$ up to T$^*$. 
The longest Fermi-arc is found in optimally doped samples \cite{nature_phys}. We measured the Fermi-arc in the pristine sample just above T$_c$; the tip of the arc is located at a Fermi-angle of 15$^o$. It is easy to see from Fig. \ref{Fig2}a that nothing special happens around that point when T$_c$ is lowered. Moreover, if there were a change of 20\% (The change in T$_c$) in the gap value around that point, we should be able to detect it easily. 
 Our data prove that there is no gap around the nodes which is proportional to T$_c$. We cannot rule out, however, a scenario in which the gap-shape changes with doping. The question of deviations from a simple d-wave gap in Bi2212 for very low doping samples remains controversial \cite{Tanaka2006,Utpal_natphys}.

\begin{figure}
\begin{center}
\includegraphics[width=8.0cm]{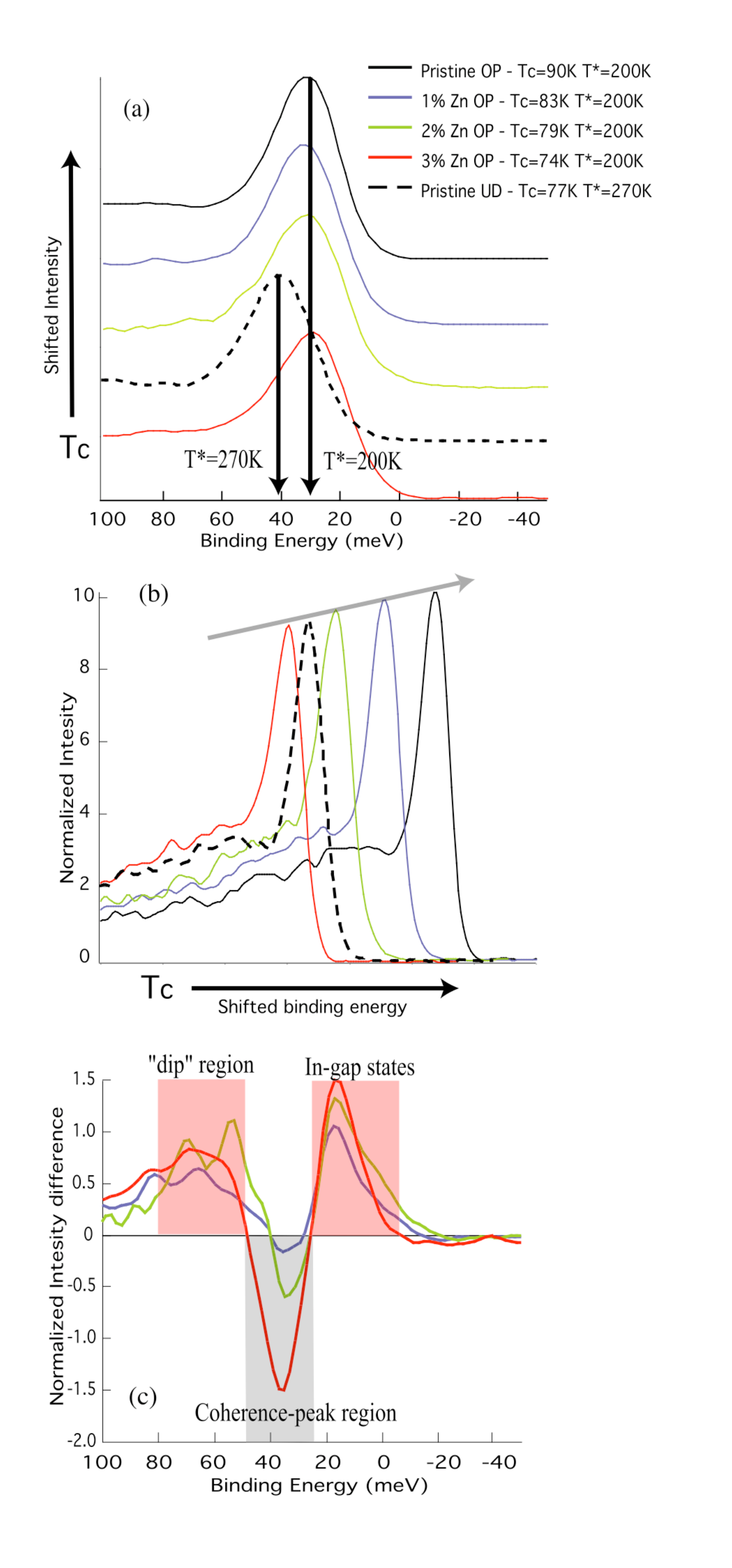}
\end{center}
\caption{
(a) The antinodal EDCs at $T=25K$. The curves are ordered according to the T$_c$ of the samples (vertical offset). All the optimally-doped samples have the same gap, the under-doped sample has a higher T$^*$ and consequently a larger gap. (b) The same spectra as in a, again ordered according to T$_c$ (Horizontal offset). The peak-height grows linearly with T$_c$, with no noticeable dependence on T$^*$. 
(c) The Anti-nodal EDC of the pristine sample was subtracted from the EDCs of the Zn substituted samples. While there is missing weight in the coherence-peak region additional sates are created in the gap and at higher energies filling the so called "dip".}
\label{edcAN}
\end{figure}

In Fig~\ref{edcAN}, the anti-nodal EDCs at 25K of all the samples are shown. The EDCs were normalized to their high binding-energy values and the background was removed by subtracting an EDC measured deep in the unoccupied region. 
This figure contains our most important findings. In panel a, we compare all the EDCs offset vertically and ordered according to their T$_c$s. The peak position, which is a measure of the gap, of all the samples having a T$^*$ of 200K (optimally doped samples) is the same regardless of their T$_c$. Only the underdoped sample has a higher T$^*$ (270K) and, as a result, a larger gap.
In panel b the same EDCs are shown, this time with an horizontal offset, again ordered according to their T$_c$. The peak height clearly increases with T$_c$, and it does not depend on T$^*$. We emphasize that all the Zn-substitued samples (solid-lines) have the same doping, so the peak-height is not simply related to the amount of charge carriers in the sample.  
In panel c, the difference between the pristine optimally doped sample and the Zn substituted samples are shown. The normalized  EDC of the pristine sample was subtracted from the normalized EDC of each of the Zn substituted samples.
The results indicate that the Zn reduces the coherence-peak weight, but creates states in other energy regions. In-gap states are created in agreement with STM data \cite{DavisSTM} and previous ARPES experiments \cite{TerashimaARC}. 
At higher binding energies there is accumulation of states that washes-out the characteristic peak-dip-hump line shape\cite{Terashima_nutaphys}.
Figure \ref{edcAN} suggests that the Zn reduces the coherence peaks at the gap-edge in a way which follows the decrease in T$_c$ and in $n_s$ in a very similar way to what is found in pristine Bi2212 where the coherence peak weight was found to be proportional to T$_c$ \cite{Feng}. 

As the beam-spot size is much larger than the average Zn spacing, we are measuring an average over regions near the impurities and regions far from the impurities. The data is consistent with the Zn imputies being  very local disturbers. If the Zn destroys SC, but its effect is very short ranged we would expect a very small effect on the average gap, since the overall volume-fraction taken by the Zn is very small. On the other hand, like in many other systems even a small amount of defects can substantially reduce the stiffness of the entire system. In the cuprates the superfluid stiffness sets T$_c$\cite{uemura}. The different response of the gap and of T$_c$ to the Zn substitution is a manifestation of the short coherence length and low superfluid density in the cuprates.

Our results show that the size of the energy gap is insensitive to variation in T$_c$, it is controlled only by T$^*$. A more complete review of the data reveals the role of T$_c$.  Taking into account that the coherence peaks appear only below T$_c$, the dramatic change in the gap-shape on crossing T$_c$ \cite{Kanigel_PRL07} and the fact that at low temperature the height of the coherence peaks follows T$_c$ suggests an unusual picture. In this picture, unlike in the conventional SC, the low temperature SC gap is not simply related to T$_c$, but T$_c$ affects other parts of the low temperature electronic spectra.

We are grateful to M. Randeria, A. Balatsky and U. Chatterjee   
for helpful discussions and to G. Koren for help with the sample preparation. 
This work was supported by the Israeli Science Foundation. 
 The Synchrotron Radiation Center is 
supported by NSF DMR-0084402. 
Part of this work
was performed at the Swiss Light Source, Paul Scherrer
Institut, Switzerland.

\end{document}